# Bond-Selective Transient Phase Microscopy

Delong Zhang†[1], Lu Lan†[1], Yeran Bai[1,2,3], Hassaan Majeed[4],

Mikhail E. Kandel[5], Gabriel Popescu[4,5*], Ji-Xin Cheng[1,6,7]*

1. Department of Biomedical Engineering, Boston University, Boston, MA 02215, USA

2. National Laboratory on High Power Laser and Physics, Shanghai 201800, China

3. Key Laboratory of High Power Laser and Physics, Shanghai Institute of Optics and Fine Mechanics, Chinese Academy of Sciences, Shanghai 201800, China

4. Department of Bioengineering, University of Illinois at Urbana-Champaign, Champaign, IL 61801, USA

5. Department of Electrical and Computer Engineering, University of Illinois at Urbana-Champaign, Champaign, IL 61801, USA

6. Department of Electrical & Computer Engineering, Boston University, Boston, MA 02215, USA

7. Photonics Center, Boston University, Boston, MA 02215 USA

†Authors contributed equally to this work

Correspondence to: jxcheng@bu.edu, gpopescu@illinois.edu

**Abstract**

Phase-contrast microscopy converts the optical phase introduced by transparent, unlabeled specimens into modulation in the intensity image. Modern phase imaging techniques are capable of quantifying phase shift at each point in the field of view, enabling non-destructive applications in materials and life sciences. However, these attractive features come with the lack of molecular information. To fulfill this gap, we developed a bond-selective transient phase (BSTP) microscope using infrared absorption to excite molecular vibration, resulting in an optical phase change detected through a customized phase microscope. By using pulsed pump and probe lasers, we realized BSTP imaging with high spectral fidelity, sub-microsecond temporal resolution, submicron spatial resolution at 50 frames per second, limited only by the camera sensor. Our approach links the missing molecular bond information to the quantitative phase, which paves a new avenue for spectroscopic imaging in biology and materials science.

Ever since Antony Van Leuwenhoek's single lens microscope in the 18$^{th}$ century [1], optical bright field microscopy has been relying on absorption as the main contrast mechanism in an intensity image. Thus, samples with little absorption or scattering, such as biological cells, generate weak intensity modulation and low contrast images. Nevertheless, transparent samples change the probing light significantly in terms of optical phase delay. By introducing an additional quadrature phase shift between the incident and scattered light, Frits Zernike was able to convert sample's path length map into intensity contrast, which allowed the investigation of transparent, unlabeled specimens [2]. For his invention of the phase-contrast microscope, Zernike received the 1953 Nobel Prize in Physics [3]. Later, *holography* was proposed as an approach for recording phase information into the detected intensity via interference between a sample and reference field [4,5]. For his invention of the holographic method, Dennis Gabor received the 1971 Nobel Prize in Physics [6]. Holography has since advanced significantly, when digital cameras and powerful computer processors became readily available [7]. The concept of phase has expanded in the field of imaging, providing the capability of capturing phase images quantitatively. Modern phase imaging can be realized using either holographic [8-15] or non-holographic approaches [16-19], and found broad applications in cellular dynamics and disease diagnosis [20-24]. Furthermore, recent developments in interferometric microscopy has demonstrated a path length sensitivity down to the sub-nanometer scale [25], promising a broader range of applications in imaging.

However, in the visible spectrum, the phase of a sample is largely insensitive to the chemical composition. Lacking such specificity makes it difficult to apply phase imaging to understand molecular interactions in a complex system. Towards this goal, integration of phase imaging with fluorescent labeling has been reported [26,27]. Yet, the fluorescent labels have their fundamental limitations, including photo-bleaching, perturbation of biological structures, and incapability of labeling small molecules. On the contrary, intrinsic molecular bond vibration can be utilized as a label-free contrast for chemical imaging, i.e. vibrational spectroscopy, in which infrared (IR) and Raman spectroscopy are the major modalities. Compared to Raman scattering, IR absorption is a relatively strong effect, observed as the attenuation of light by the sample. An IR spectroscopy database of common chemicals was published [28] as early as 1905. However, because of the long wavelength compared to the visible spectrum, the diffraction limit made IR imaging poor in spatial resolution. Moreover, IR spectroscopy measures the total loss of input beam, which is difficult to extract the intrinsic absorption property from other attenuation effects, such as sample scattering and reflection. These disadvantages limited the adoption of IR imaging. Nevertheless, the energy from IR absorption causes a temperature increase in the specimen, which changes the refractive index via thermo-optic effect, resulting in a change in the optical path length of the specimen. Such change can be measured and quantified by phase imaging approaches, providing the intrinsic molecular spectroscopy of the specimen.

Base on this idea, we present a bond-selective transient phase (BSTP) microscope that brings chemical information to phase imaging through infrared light perturbation. The temperature increase caused by IR absorption is transient, in which the heat dissipates in as fast as a few microseconds. Therefore, a high-temporal resolution phase imaging system is required. However, most current phase imaging apparatus, with imaging speed of up to a few thousand frames per second [25,29], are still insufficient to record such transient phase change. To address this challenge, we developed a wide-field pump-probe imaging platform offering sub-microsecond temporal resolution. Specifically, by utilizing a sub-microsecond burst of laser pulses to *probe* the transient phase change caused by a nanosecond pulsed mid-infrared *pump* laser, we achieved widefield BSTP imaging with sub-microsecond temporal resolution.

The principle of BSTP imaging is illustrated in **Fig. 1**. First, using a common-path, off-axis diffraction phase microscopy [12], we generate the quantitative phase image of the unperturbed sample, which is referred to as the "cold" frame. Next, mid-infrared pulses illuminate the sample, generating absorption which, as a result, induces refractive index changes and a modified quantitative phase image, referred to as the "hot" frame. The phase difference between the hot and cold frames is directly proportional to the IR absorption at the sample, which spectroscopy can be acquired by scanning the IR pump wavelength.

The optical phase shift, $\varphi$, can be described with regard to the refractive index $n$ and the thickness $l$ of an object in air, as $\varphi = \frac{2\pi}{\lambda}(n-1)l$. Mid-IR pulses with frequency $\omega$, energy $E$, and illumination area $A$ at the sample are absorbed according to the vibrational absorption coefficient $\mu(\omega)$, causing a local temperature increase $\Delta T$. For simplicity, we assume the single pulse absorption a steady-state adiabatic process, considering first order Taylor expansion of the Lambert-Beer law, namely, $e^{-\mu(\omega)l} \approx 1-\mu(\omega)l$. Thus, the estimated temperature change, $\Delta T$, can be obtained as

$$\Delta T = \frac{Q}{C_p m} = \frac{(1-e^{-\mu(\omega)l})E}{C_p \rho A l} \approx \frac{\mu(\omega)E}{C_p \rho A} \tag{1}$$

where $C_p$ is the specific heat, and $\rho$ is the mass density. This local temperature rise results in a change in the refractive index,

$$\Delta n = \frac{dn}{dT}\Delta T = \alpha \Delta T, \tag{2}$$

and a change in thickness,

$$\Delta l = (\frac{1}{l}\frac{dl}{dT})l\Delta T = \beta l\Delta T. \tag{3}$$

In Eqs. 2-3, $\alpha$ is the thermo-optic coefficient and $\beta$ is the linear thermal expansion coefficient. The measured optical phase change, $\Delta\varphi$, can be obtained by finding the difference between the hot and cold phase frames. For small $\Delta n$ and $\Delta l$, we obtain

$$\Delta\varphi = \varphi_{hot} - \varphi_{cold} = \frac{2\pi}{\lambda}[(n+\Delta n)(l+\Delta l) - nl] = \frac{2\pi}{\lambda}(\Delta nl + n\Delta l). \tag{4}$$

Plugging Eqs. (1)-(3) into Eq. 4, we obtain

$$\Delta\varphi(\omega) = \gamma l \frac{E}{A\lambda}\mu(\omega), \tag{5}$$

where $\gamma = \frac{2\pi(\alpha + n\beta)}{C_p\rho}$.

In Eq. 5, three factors have distinct meanings: $\gamma l$ is the physical property of the sample, $\frac{E}{A\lambda}$ is the pump pulse property, and $\mu(\omega)$ is the IR spectroscopic absorption. It is clear that $\Delta\varphi$ is quantitatively related to the chemical content of the sample. Spectroscopic imaging could thus be obtained by tuning the wavelength of a narrow bandwidth IR laser source, providing molecular information to the quantitative phase images.

Unlike conventional phase imaging, BSTP imaging requires microsecond-scale temporal resolution to detect the transient phase change. This process is fast because the absorption induced heat dissipates immediately after an IR pulse arrives at the sample. Depending on the sample properties, the thermal decay constant ranges from a few microseconds to hundreds of microseconds. Therefore, microsecond level temporal resolution is required. However, commercially available cameras are not capable of recording a million frames per second at a pixel resolution sufficient for an acceptable interferogram. To address this challenge, instead of the commonly used continuous wave light source, we adopted a pulsed probe laser to construct a time-gated scheme, in which all photons to the camera are from the short duration of the probe pulse. Moreover, high temporal resolution thermal decay can be recorded by adjusting the time delay between the probe pulse and the heating pulse, providing important thermodynamics of the specimen.

Based on this idea, we built a BSTP imaging system capable of detecting a local phase change induced by a transient IR absorption (**Fig. 2a**). The temporal resolution of the system is determined by the probe laser pulse width. The probe beam is from an 80-MHz femtosecond pulsed laser and is chopped by an acousto-optic modulator (AOM) to an arbitrary burst duration down to 70 ns. An important feature of femtosecond pulses for diffraction phase microscopy is the much shorter coherence length, which suppresses much of the speckle noise present when using monochromatic light. A transmission grating was positioned at the conjugate plane of the sample to split the probe beam into multiple orders. Of these we block all but two: the first order, which is filtered by a pinhole at the Fourier plane, and the second order, which passes unaltered [30]. The two waves interfere, the first order acting as a reference, and generate a stable interferogram at the camera sensor plane. A pulsed IR laser illuminates the field of view to generate a hot frame. The hot and cold frames are compared to generate a BSTP image. Because our camera integrates over 10 ms exposure times, we send 9 pump-probe pulse pairs for each camera exposure, such that we maximize signal to noise ratio (**Fig. 2b**). The chopper modulates the 150 kHz IR pulse train into small bursts, down to single pulses per burst. The shutter closes the IR beam when exposing cold frames. The timing electronics is based on the IR laser repetition rate, in which frequency dividers were set to generate clocks for the chopper, AOM, shutter, and camera (**Fig. 2c**).

We measured the temporal profile of the signal from an oil film by scanning the delay of the probe pulse delay. Because the camera only records photons from the probe pulse, the temporal resolution of the system is determined by the probe pulse width. The probe width was set at 900 ns, which is a temporal resolution that, in traditional detection schemes, would only be achievable by a camera acquisition rate of 1.1 million frames per second, significantly above what is commercially available. By shining IR pulses tuned to 2950 cm$^{-1}$ for the CH stretching vibration with pulse energy of 0.67 $\mu J$ over a 32-$\mu m$ diameter spot at sample, BSTP images were recorded at each time delay that was tuned electronically by the pulse generator. Note that differential phase images were performed by subtracting hot phase images out of cold phase images for better visualization. A representative BSTP image is shown in **Fig. 3a**; the temporal profile, i.e. phase shift vs. time delay is shown in **Fig. 3b**. Because of the 900 ns probe pulse width, individual IR pulses were clearly resolved for the six pulses with a 6.6 $\mu s$ interval. The maximum signal was observed at the end of the IR pulses, followed by a thermal decay with an exponential decay constant $\tau = 130.8 \mu s$, which was the slowest decay in our observations. Smaller objects would have much shorter decay constant. To allow sufficient cooling time, 1 ms interval was set between IR pulse bursts. In the following experiments, the probe pulse was set immediately after the last IR pulse, where maximum signal was acquired.

To verify that the measured phase change $\Delta\varphi$ is proportional to the absorption coefficient $\mu(\omega)$, we scanned the IR wavelength to acquire BSTP images and compared the profile to that from Fourier-transform infrared (FTIR) spectroscopy (**Fig. 3c**). The $\Delta\varphi$ values from each image were plotted along the spectrum of the same sample measured on a standard FTIR microscope, showing good agreement. Furthermore, by fitting the 2910 cm$^{-1}$ peak, a FWHM 8.9 cm$^{-1}$ was obtained, which is consistent with the spectral width of the IR laser of between 6 and 9 cm$^{-1}$. It is clear that BSTP imaging is capable of generating high fidelity spectroscopic images.

BSTP imaging involves an IR (pump) excitation pulse and a green probe pulse and can be described as a pump-probe process. It is obvious that the signal is proportional to the pump power, as validated by experiments (**Fig. 3d**). The signal is, however, independent of the probe power, proving that the probe pulse-sample interaction is linear (**Fig. 3e**). This is because the probe beam is not absorbed significantly by the sample. Nevertheless, higher probe power increases the signal-to-noise ratio (SNR) by reducing the effect of shot noise in the phase image. Further noise analysis in BSTP revealed the relation between SNR and number of photons $N$ to be $SNR \propto N^{0.39}$ (**Fig. 3f**). Ideally, for shot-noise limited signal, the power term would be 0.5, according to the Poisson distribution. This faster saturation of SNR is likely due to mechanical noise surviving in our common path interferometer. The SNR can be further improved by averaging over a larger number of pulses for each recording, while shortening the exposure time by using a faster camera. Nevertheless, this proof of principle BSTP imaging system is already sensitive to the mrad scale of IR-absorption-induced phase shifts.

To demonstrate the chemical selectivity, we performed BSTP imaging on polyurethane (PU) beads, providing spectroscopic information at each pixel to augment the conventional phase image (**Fig. 4**). The beads sample (U7-D50, HOS-Technik GmbH, Austria) has a size distribution with 50% of particles at 7 $\mu m$. BSTP imaging showed good contrast when the IR laser was tuned to the absorption peak at 2980 cm$^{-1}$ (**Fig. 4b**), while minimal contrast was observed at an off-resonance frequency of 2700 cm$^{-1}$ (**Fig. 4c**). We further scanned the pump laser and generated a phase-based IR spectrum of the sample, compared it to the FTIR absorption profile of the same sample (**Fig. 4d**). Note that the FTIR microscope had difficulty acquiring spectrum from single beads due to the IR diffraction limit; it also suffers from sample diffraction artifacts. Therefore, the FTIR spectrum showed an obvious baseline. On the contrary, BSTP imaging provides a baseline-free spectrum, because the signal only rises from the actual IR absorption. As a result, the small characteristic peaks at 2900 cm$^{-1}$ and 2980 cm$^{-1}$ were observed in $\Delta\varphi$, but not obvious in FTIR measurements. We also compared the spatial resolution using a small object in the image. **Fig. 4e** compares the line profiles of the bead between the raw phase and the BSTP images, showing consistency in between. It is also worth noting that the BSTP images were taken with single IR pulses at 100 frames per second, i.e. 50 pairs of hot and cold frames, with a 30-frame average, resulting in a speed of 1.67 images per second. Furthermore, the chemical selectivity of BSTP imaging was demonstrated at an interface between two chemicals, dimethyl sulfoxide (DMSO) and olive oil. In conventional phase imaging, although a contrast could be observed, it is virtually impossible to identify the molecular bonds. On the contrary, BSTP imaging revealed the IR spectroscopy response at each pixel, providing new insights on molecular structures that are beyond the reach of conventional phase imaging.

The mechanism of BSTP imaging is not limited to the mid-IR region. The excitation wavelength can be extended to the near-IR window for overtone spectroscopy for a wider range of applications. As an example, the overtone absorption of 10-$\mu m$ Poly-methyl-methacrylate (PMMA) beads provides contrast for BSTP imaging. Collectively, BSTP imaging combines the power of phase imaging with spectroscopy insights, enabling new applications for chemical imaging.

We note that BSTP imaging detects local changes of phase shifts induced by mid-infrared absorption of chemical bonds inside a specimen. The local temperature increase modifies both the physical length and the refractive index of the sample, which contribute together to the optical phase. While the physical dimension always increases after IR absorption, the refractive index change can be either positive or negative, depending on the material properties. For example, PMMA has a linear thermal expansion coefficient of $75 \times 10^{-6} K^{-1}$, but a negative thermo-optic coefficient [31] of $-130 \times 10^{-6} K^{-1}$. In BSTP imaging, these two terms result in a rather small phase change coefficient, $-18 \times 10^{-6} K^{-1}$. Therefore, if the physical dimension change and the refractive index change could be recorded separately, the net coefficient will be $224 \times 10^{-6} K^{-1}$, which promises approximately 12 times increase of signal level. Thus, optical phase tomography techniques [32-34], which spatially decouple the refractive index and the thickness of the sample, can be used to further enhance our BSTP signals, especially when the thermo-optic coefficient is negative.

BSTP imaging, as a widefield imaging platform, has the speed advantage over 3 orders of magnitude faster than the recently developed mid-infrared photothermal (MIP) microscopy, which utilizes a visible beam to probe the photothermal effect caused by IR absorption in a point-scanning manner [35-39]. MIP microscopy provides depth-resolution, sub-micrometer lateral resolution, and micro-molar detection sensitivity. MIP has enabled many applications in biology and materials science, including high-resolution chemical imaging of live cell, live organism imaging, cellular metabolites, drug molecules in cells and products, and energy materials [35,39-41]. With the much-improved speed of chemical imaging, BSTP imaging will enable a much broader applications in the related fields.

In summary, we demonstrated a BSTP imaging platform that provides the spectral and temporal dynamics of a specimen with high spectral fidelity and sub-microsecond temporal resolution. We reached a high speed of 50 images per second, limited by the camera. Though we demonstrated the proof-of-concept using a diffraction phase microscopy frame, our pump-probe approach can be applied to add chemical information to phase imaging using non-interferometric methods. By unveiling the chemical composition and thermodynamics information in addition to conventional quantitative phase imaging, BSTP imaging promises wide applications in biology and materials science.

**Acknowledgments**

The authors thank Yimin Huang for helping to prepare the polymeric phantoms. The project is supported by a R01 Grant GM126049 to J.X.C and the National Science Foundation grant CBET-0939511 STC (to G.P.).

**Author Contributions**

D.Z., G.P., and J.X.C conceived the idea; D.Z. and L.L. designed the imaging system and carried out the implementation with the input from H.M., M.K., Y.B., G.P., and J.X.C.; L.L. and D.Z. collected and analyzed the data; D.Z. and L.L. wrote the manuscript supervised by J.X.C.; All authors discussed the results and contributed to the final manuscript.

**Competing Interests**

The authors declare no competing interests.

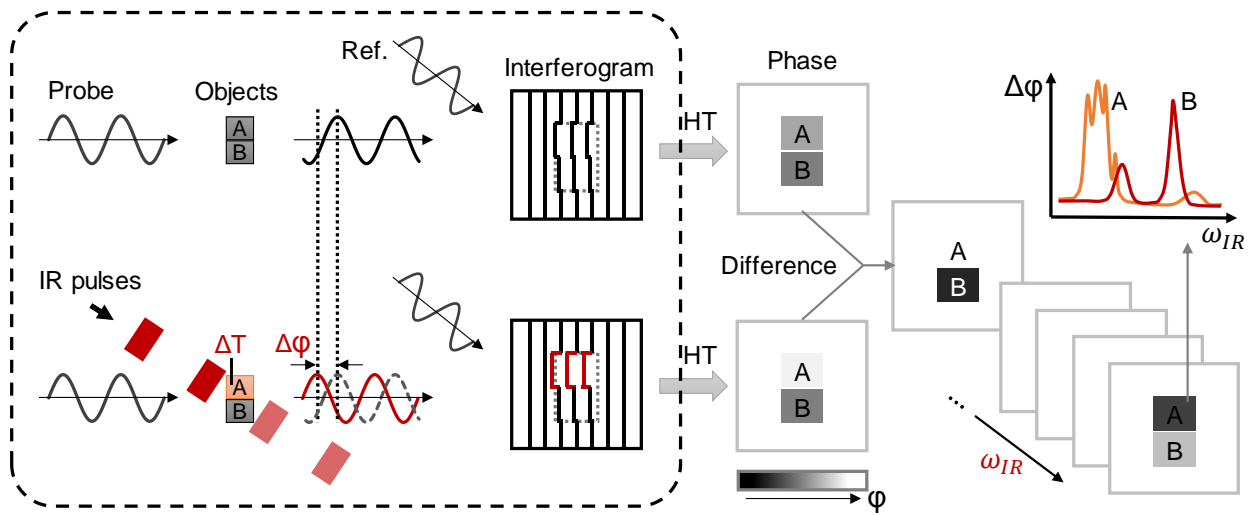

**Figure 1. Principle of BSTP imaging.** The probe light passes through the sample and generated an interferogram at the camera with the reference beam. A phase image of the sample is then retrieved via Hilbert transform (HT). Next, IR pulses at the sample induces vibrational absorption thus a local temperature rise ΔT that changes the local optical phase. The transient change of phase $\Delta\varphi$ can then be obtained by subtraction between the adjacent phase images. By tuning the excitation frequency $\omega_{IR}$, the spectroscopy at each pixel can be acquired.

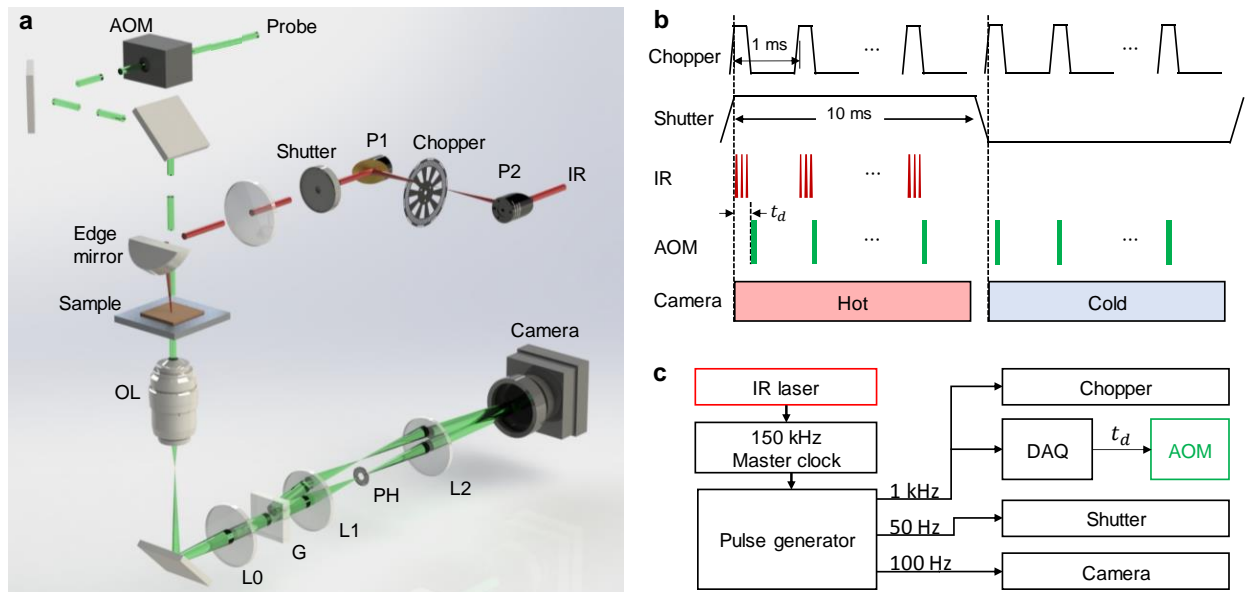

**Figure 2. Scheme of BSTP microscope. a**, Illustration of the BSTP microscope. A common-path diffraction phase microscope was built on an inverted microscope, and an IR beam was introduced for the BSTP imaging. AOM: acousto-optical modulator. P1, P2: gold parabolic mirrors. OL: objective lens. L0, L1, L2: lenses. G: grating. PH: pinhole. **b**, Timing diagram. $t_d$: delay of the probe pulse relative to the start of the IR pulses in each burst. **c**, Clock hierarchy.

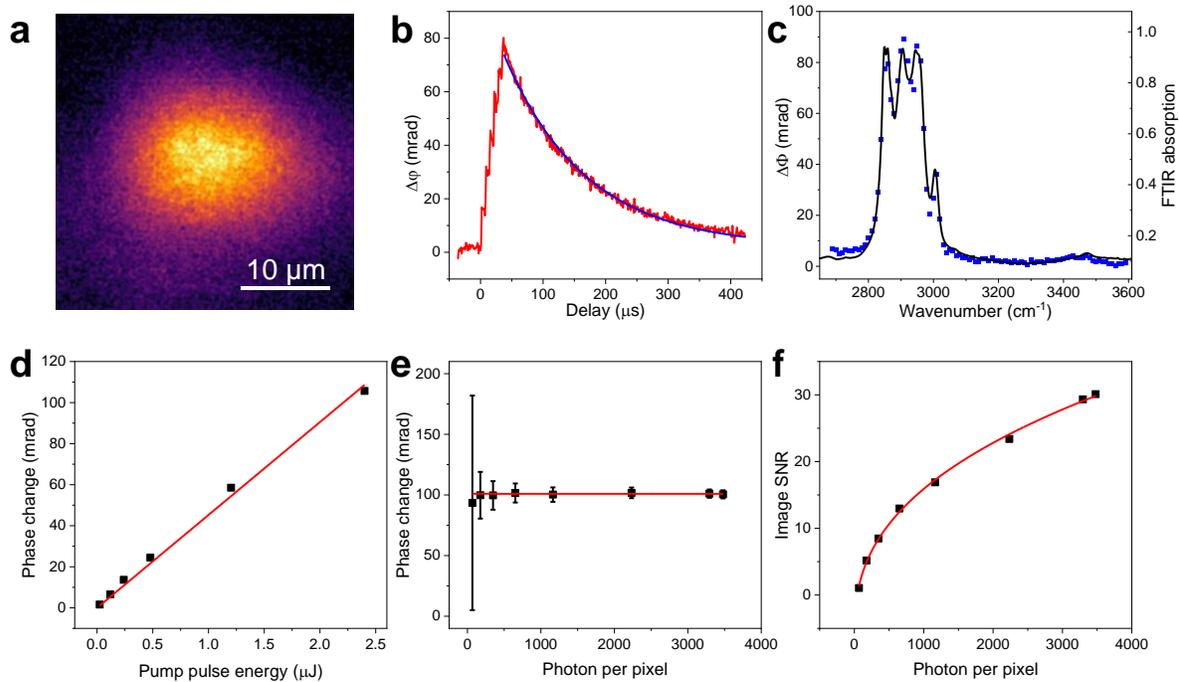

**Figure 3. Characterization of BSTP imaging. a**, BSTP imaging of an oil film sample. There were 6 IR pulses with 6.67 $\mu s$ interval for each image. The IR wavelength was at 2950 cm$^{-1}$. **b**, Temporal profile of BSTP signal. **c**, Spectral fidelity of BSTP signal (squares), as compared to the standard FTIR spectrum (line). **d**, Pump power dependence. BSTP signals at different IR power (squares) and the linear fitting curve (line). **e**, Probe power dependence. BSTP signals with different probe power (squares) and the linear fit (line). **f**, Signal-to-noise ratio of BSTP signal with different probe power (squares) and the fitted curve (line).

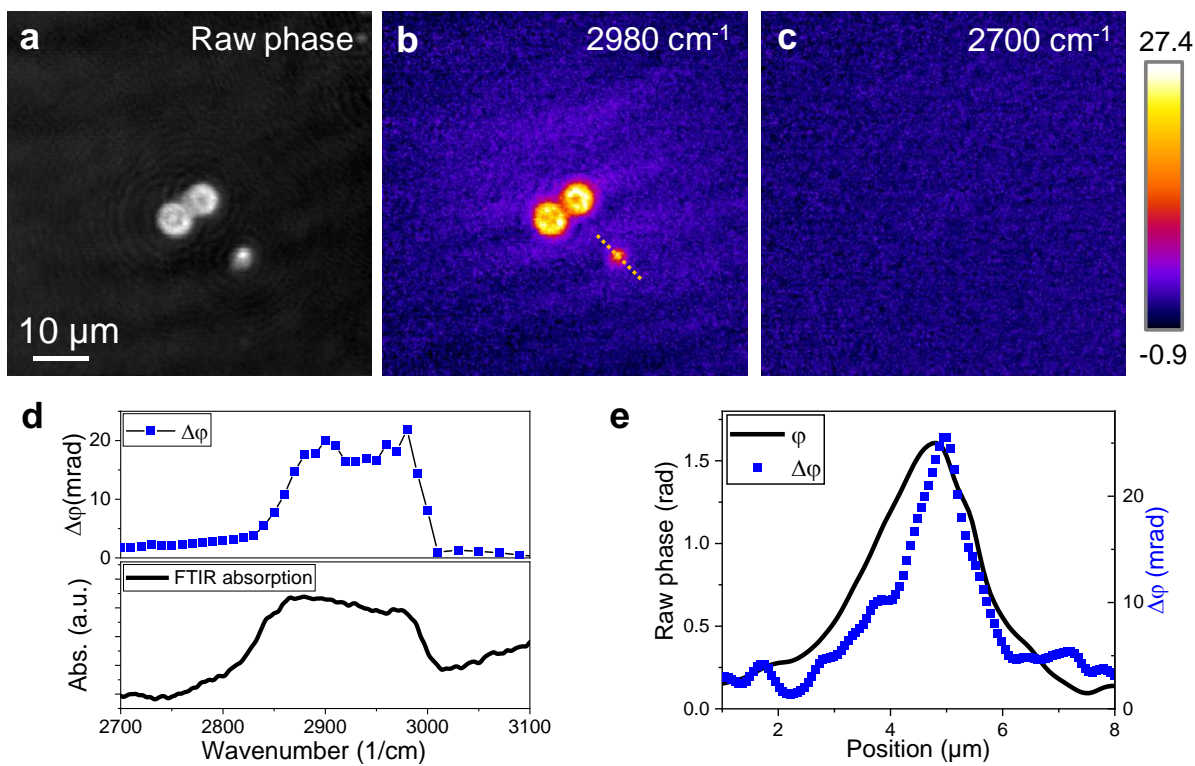

**Figure 4. BSTP imaging of polyurethane beads. a**, Quantitative phase image of PU beads. **b,c**, BSTP imaging of PU beads at its absorption peak of 2980 cm$^{-1}$ and off-resonance peak at 2700 cm$^{-1}$, respectively. **d**, BSTP spectral profile of PU beads (top) and the measured standard FTIR spectrum (bottom). **e**, BSTP (blue squares) and raw phase (black line) profiles of the smaller PU bead across the dashed line marked in **b**.

18